\documentclass[11pt,reprint,showkeys, amsmath,amssymb, aps, nofootinbib, floatfix]{revtex4-2}
\DeclareUnicodeCharacter{2212}{\textendash}
\usepackage{graphicx,subfigure}
\graphicspath{{figures/}}
\usepackage[export]{adjustbox}
\usepackage{dcolumn}
\usepackage[outdir=./epsTopdf/]{epstopdf}
\usepackage{bm}
\usepackage{xcolor}
\usepackage{slashed,cancel}
\usepackage{multirow}
\usepackage{float}
\usepackage{capt-of}
\usepackage{comment}
\usepackage{mathrsfs}
\usepackage{enumitem}
\usepackage[colorlinks=true,linkcolor=blue,urlcolor=blue,citecolor=blue]{hyperref}

\allowdisplaybreaks
\begin{document}

\title{Probing CP-violating Higgs-Gauge couplings with Higgsstrahlung at $e^-e^+$ collider}

\author{Amir Subba}
\email{amirsubba@iitg.ac.in}
\author{Subhaditya Bhattacharya}
\email{subhab@iitg.ac.in}
\author{Abhik Sarkar}
\email{sarkar.abhik@iitg.ac.in}
\affiliation{Department of Physics, Indian Institute of Technology Guwahati, Guwahati, Assam, 781039, India}

\begin{abstract}
We investigate the sensitivity of a future high-luminosity $e^-e^+$ collider operating at $\sqrt{s}=250~\text{GeV}$ to CP-violating and CP-conserving anomalous $hVV$ interactions via the Higgsstrahlung process. The effects of new physics are parameterized in the Standard Model Effective Field Theory~(SMEFT) framework through six dimension-6 operators modifying the $hVV$ vertices. Using polarized beams and exploiting polarization and spin correlation asymmetries reconstructed from Higgs decay products, we perform a comprehensive analysis across the three dominant decay modes, $h\to b\bar{b}$, $WW^\star$, and $ZZ^\star$. The $h\to WW^\star$ channel exhibits the highest sensitivity to $\mathcal{O}_{HW}$ and $\mathcal{O}_{H\widetilde{W}}$, while the $b\bar{b}$ mode constrains the remaining operators with high statistical precision. Sensitivity studies incorporating luminosity scaling and systematic uncertainties show that the projected bounds improve significantly with increasing integrated luminosity, but saturate once experimental systematics exceed the few-percent level. These results highlight the crucial role of spin-based observables and beam polarization in achieving sub-percent precision on SMEFT coefficients at future lepton colliders.
\end{abstract}

\maketitle

\section{Introduction}
\label{sec:intro}
It has been over a decade since the discovery of the Higgs boson at the Large Hadron Collider~\cite{ATLAS:2012yve,CMS:2012qbp,CMS:2013btf}, which provided the first suggestive evidence for the mechanism of Electroweak Symmetry Breaking~(EWSB)~\cite{Englert:1964et,Guralnik:1964eu,Higgs:1964pj}, a cornerstone of the Standard Model~(SM). Since then, considerable effort has been devoted to studying its properties, including its mass, electric charge, CP nature, and its interaction strengths with other SM particles. These measurements have, so far, shown good agreement with the SM predictions. However, the current experimental precision still allows for potential deviations that could signal physics beyond the Standard Model~(BSM). As the Higgs couplings to SM particles are proportional to their masses for fermions and to the square of their masses for massive gauge bosons, precise measurements of these couplings also serve as a crucial test of the mass generation mechanism.

Among the various Higgs couplings, those involving massive gauge bosons are particularly significant. They play a central role in several Higgs production mechanisms, such as vector boson fusion (VBF) and Higgs-strahlung ($Vh,V\in \{W,Z\}$), which are subdominant only to the gluon fusion process at hadron colliders and constitute the dominant production channels at lepton colliders. Furthermore, the second most prominent Higgs decay mode, $h \to WW^\star$, directly involves this coupling, further emphasizing its importance. Therefore, a precise measurement of the $hVV$ coupling, both within the SM and in the presence of possible BSM contributions, is crucial for probing the properties of the Higgs boson.

Numerous BSM scenarios have been proposed to address theoretical and experimental issues that SM alone can't handle. However, no such evidence has been experimentally gathered yet. In the absence of an established BSM framework, the effective field theory (EFT) approach has emerged as a model-independent tool to parameterize potential BSM effects through higher-dimensional operators built from known SM fields. Among these, the Standard Model Effective Field Theory~(SMEFT) has gained significant traction, particularly for collider studies, as it incorporates BSM effects via operators that are invariant under the SM gauge symmetries.

The $hVV$ coupling has been widely studied in both specific BSM models and within the SMEFT framework~\cite{Godbole:2007cn,Godbole:2013saa,Godbole:2014cfa,He:2019kgh,Nakamura:2017ihk,Banerjee:2019pks,Rao:2020hel,Bizon:2021rww,Corbett:2012dm,Corbett:2012ja,Bhattacharya:2024sxl}, especially in the context of the LHC. However, factors such as substantial QCD backgrounds, the variable sub-process center-of-mass energy ($\sqrt{\hat{s}}$), and the lack of polarization in the initial states make it challenging to perform precision studies of electroweak couplings like $hVV$ at hadron colliders. On the other hand, lepton colliders offer a much cleaner environment for studying electroweak physics, characterized by reduced QCD backgrounds, fixed $\sqrt{s}$, and the possibility of initial-state beam polarization. These features significantly enhance the precision with which electroweak couplings can be measured. In particular, the $hVV$ couplings have been extensively studied in the context of lepton colliders through both the $Vh$, VBF and multi-gauge processes~\cite{Hagiwara:1993sw,Hagiwara:2000tk,Biswal:2005fh,Biswal:2008tg,Biswal:2009ar,Rindani:2010pi,Rindani:2009pb,Anderson:2013afp,Craig:2015wwr,Beneke:2014sba,Khanpour:2017cfq,Zagoskin:2018wdo,Li:2019evl,Chiu:2017yrx,Durieux:2017rsg,Subba:2024aut,Subba:2025pos,Atwal:2025yhw,Ruhdorfer:2024dgz,Ogawa:2017bmg}.

In the current work, we investigate anomalous $hZZ, hZ\gamma$ and $hWW$ couplings via $e^-e^+ \to Zh$ production, within the framework of SMEFT with initial beam polarization set to the baseline parameters of upcoming International Linear Collider (ILC)~\cite{ILC:2007bjz,ILC:2007oiw,ILC:2007vrf,ILC:2013jhg,Adolphsen:2013kya} with $\sqrt{s}=250$~GeV. The SMEFT-induced anomalous contributions include both CP conserving and CP violating Lorentz structures, which will be detailed in the following section. In~\cite{Bhattacharya:2025jhs}, we explored these couplings at the $e^-e^+$ collider setup using the Optimal Observable Technique~(OOT), which determines the maximal sensitivity achievable from cross section-based observables. While that approach demonstrated significant improvement in constraining the CP even operators beyond current LHC bounds, the sensitivity to CP odd operators remained limited. This is primarily because such operators do not interfere with the SM amplitudes in total cross-section-based measurements, making their effects harder to detect through traditional observables.

Polarization and spin-correlation observables provide a more incisive probe of CP-odd operators, as they are sensitive to interference effects that remain inaccessible in inclusive cross-section measurements. The polarization of the $Z$ boson in the Higgsstrahlung process has previously been explored in Refs.~\cite{Rao:2020hel,Biswal:2005fh,Rindani:2009pb,Nakamura:2017ihk} and related works. In the present study, we extend this approach by simultaneously analyzing the $Z$-boson polarization and the spin correlations among multiple vector bosons originating from the Higgs decays. This enables a comprehensive assessment of the sensitivity to anomalous $hVV$ interactions. Specifically, we consider three dominant Higgs decay channels: $h \to b\overline{b}$, $h \to WW^*$, and $h \to ZZ^*$. While the $b\overline{b}$ mode provides the highest event statistics, the $WW^*$ and $ZZ^*$ channels offer complementary advantages by introducing anomalous contributions at both the production and decay vertices. The resulting spin-correlated observables constitute a rich set of kinematic handles, whose combined effects substantially enhance the overall sensitivity to both CP-even and CP-odd operators.

The paper is organized as follows. Section~\ref{sec:smeft} introduces the SMEFT framework employed in this work. The reconstruction methodology of spin related observables, which serve as sensitive probes of CP-violating effects, are detailed in Sec.~\ref{sec:spin}. The results of the sensitivity analysis across the three principal Higgs decay channels, with particular emphasis on CP-odd operators, are presented in Sec.~\ref{sec:res1}. A global interpretation combining information from all decay modes is provided in Sec.~\ref{sec:res2}. Finally, the main conclusions and outlook of the study are summarized in Sec.~\ref{sec:con}.

\section{SMEFT Framework}
\label{sec:smeft}
In the absence of direct observation of new physics~(NP) at current experimental energies, the effective field theory (EFT) framework emerges as one of the most promising approaches to study effects of NP. In the context of collider experiments, the effects of on-shell NP states that lie beyond the current energy reach can be indirectly probed through gauge-invariant higher-dimensional operators constructed from the SM fields. Keeping the gauge symmetry of the SM intact, the extension is formalized in the framework known as the SMEFT~\cite{Buchmuller:1985jz}. The SMEFT Lagrangian is written as
\begin{equation}
    \mathscr{L}_{\text{SMEFT}} = \mathscr{L}_{\text{SM}} + \sum_i \frac{C_i^{(5)}}{\Lambda}\mathscr{O}_i^{(5)} + \sum_{i} \frac{C_i^{(6)}}{\Lambda^2} \mathscr{O}_i^{(6)} + \mathscr{O}\left(\frac{1}{\Lambda^4}\right),
\end{equation}
where $\mathcal{O}_i$ are set of operators at dimension-$d$ constructed with the SM fields, $C_i$ are Wilson coefficients, and $\Lambda$ denotes the NP scale where decoupling occurs. Throughout the analysis, we fix $\Lambda = 1$ TeV for simplicity\footnote{The value of $\Lambda  > \sqrt{s}$ ensures the validation of EFT framework.} as cross~section for any $\Lambda$ can be obtained by appropriately rescaling. Since at each order of $d$, the amplitude becomes suppressed as $(E/\Lambda)^{(d-4)}$, the dominant SMEFT contribution comes from lowest order of dimension. Assuming the lepton and baryon number conservation, the lowest order of SMEFT begins at dimesion-$6$. We consider three CP even and their dual CP odd operators inducing anomalous contribution to Higgsstrahlung process. The relevant operators in Warsaw basis are~\cite{Grzadkowski:2010es}
    \begin{equation}
    \label{eq:op}
        \begin{aligned}
            \mathscr{O}_{H W} ~&= ~(H^{\dagger} H) W^{i}_{\mu \nu} W^{i \mu \nu},\\ 
            \mathscr{O}_{H\widetilde{W}} ~&= ~(H^{\dagger} H) \widetilde{W}^{i}_{\mu \nu} W^{i \mu \nu}, \\
            \mathscr{O}_{H WB} ~&= ~(H^{\dagger} \tau^{i} H) W^{i}_{\mu \nu} B^{\mu \nu},\\  \mathscr{O}_{H\widetilde{W}B} ~&= ~(H^{\dagger} \tau^{i} H) \widetilde{W}^{i}_{\mu \nu} B^{\mu \nu} \\
            \mathscr{O}_{H B} ~&= ~(H^{\dagger} H) B_{\mu \nu} B^{\mu \nu},\\
            \mathscr{O}_{H\widetilde{B}} ~&=~ (H^{\dagger} H) \widetilde{B}_{\mu \nu} B^{\mu \nu} 
        \end{aligned}
    \end{equation}
Here, the operators $\mathscr{O} \in \{\mathscr{O}_{H W},\mathscr{O}_{H B},\mathscr{O}_{H WB}\}$ are CP-even while the remaining subset of operators $\mathscr{O} \in \{\mathscr{O}_{H\widetilde{W}},\mathscr{O}_{H\widetilde{B}},\mathscr{O}_{H\widetilde{W}B}\}$ involving dual tensors are CP-odd. The field tensor are defined as $W_{\mu \nu}^{i} = \partial_{\mu} W_{\nu}^{i} - \partial_{\nu} W_{\mu}^{i} + g \epsilon^{ijk} W_{\mu}^{j} W_{\nu}^{k}$, $B_{\mu \nu} = \partial_{\mu} B_{\nu} - \partial_{\nu} B_{\mu}$ and the dual field tensor is $\widetilde{V_{\mu \nu}} = \epsilon_{\mu \nu \rho \sigma} V^{\rho \sigma}$ ($V = W^{i}, B$), with the completely anti-symmetric Levi-Civita tensor following standard notation. Here, $H$ is the SM Higgs doublet. After electroweak symmetry breaking, these operators induces anomalous contribution to $hZZ$ couplings of the form
\begin{align}
    \delta\, \Gamma_{hZZ} &= \kappa_{hZZ} \left(\frac{h}{v}\, Z_{\mu \nu} Z^{\mu \nu} \right) + \kappa_{h\widetilde{Z}Z} \left(\frac{h}{v}\, \widetilde{Z}_{\mu \nu} Z^{\mu \nu} \right),
\end{align}
where
\begin{align}
\label{eq:hzz}
    \kappa_{hZZ} &= \frac{2v^2}{\Lambda^2} \left[\cos^{2}{\theta_{W}}\,C_{H W} + \cos{\theta_{W}}\sin{\theta_{W}}\,C_{H WB} \right.\nonumber\\&+\left. \sin^{2}{\theta_{W}}\,C_{H B}\right],\nonumber\\
    \kappa_{h\widetilde{Z}Z} &= \frac{2v^2}{\Lambda^2} \left[\cos^{2}{\theta_{W}}\,C_{H \widetilde{W}} + \cos{\theta_{W}}\sin{\theta_{W}}\,C_{H \widetilde{W}B} \right.\nonumber\\&+\left. \sin^{2}{\theta_{W}}\,C_{H \widetilde{B}}\right].
\end{align}
 Apart from anomalous $hZZ$ coupling, the operators allows for tree level $s$-channel $Zh$ production mediated by massless photon, which otherwise is a loop process in the SM. The anomalous $hZ\gamma$ couplings is obtained to be
\begin{align}
    \delta\, \Gamma_{hZ\gamma} = \kappa_{hZ\gamma} \left(\frac{h}{v}\, Z_{\mu \nu} A^{\mu \nu} \right) + \kappa_{h\widetilde{Z}\gamma} \left(\frac{h}{v}\, \widetilde{Z}_{\mu \nu} A^{\mu \nu} \right),
\end{align}
where
    \begin{align}
    \label{eq:hza}
    \kappa_{hZ\gamma} &= \frac{v^2}{\Lambda^2} \left[2\cos{\theta_{W}}\sin{\theta_{W}}\,(C_{H W} - C_{H B}) \right.\nonumber\\&+\left. (\sin^{2}{\theta_{W}} - \cos^{2}{\theta_{W}})\,C_{H WB}\right],\nonumber\\
    \kappa_{h\widetilde{Z}\gamma} &= \frac{v^2}{\Lambda^2} \left[2\cos{\theta_{W}}\sin{\theta_{W}}\,(C_{H \widetilde{W}} - C_{H \widetilde{B}}) \right.\nonumber\\&+\left. (\sin^{2}{\theta_{W}} - \cos^{2}{\theta_{W}})\,C_{H \widetilde{W}B}\right].
    \end{align}
The inclusion of EFT also contributes to the Higgs decay to $WW^*$ via anomalous $hWW$ couplings of the form
\begin{align}
    \delta\, \Gamma_{hWW} &= \kappa_{hWW} \left(\frac{h}{v}\, W^{+}_{\mu \nu} W^{-\mu \nu} \right) \nonumber\\&+ \kappa_{h\widetilde{W}W} \left(\frac{h}{v}\, \widetilde{W}^{+}_{\mu \nu} W^{-\mu \nu} \right),
\end{align}
where
    \begin{align}
    \label{eq:hww}
    \kappa_{hWW} &= \frac{2v^2}{\Lambda^2} C_{HW}\,,\\
    \kappa_{h\widetilde{W}W} &= \frac{2v^2}{\Lambda^2} C_{H\widetilde{W}}\,.
    \end{align}

Non observation of any excess beyond SM sets the current experimental limits at $68\%$ confidence level~(CL) on the WCs of the above operators as provided by CMS~\cite{CMS:2024bua}
\begin{equation} \label{eq:lhc}
\begin{split}
    C_{H W} = [-0.79,+0.51], \hspace{0.6cm} C_{H \widetilde{W}} &= [-0.76,+0.41],  \\
    C_{H WB} = [-1.62,+1.50], \hspace{0.4cm} C_{H \widetilde{W}B} &= [-1.57,+0.83],\\
    C_{H B } = [-0.23,+0.16], \hspace{0.7cm}
    C_{H \widetilde{B}} &= [-0.23,+0.12].\\
\end{split}
\end{equation}
The anomalous couplings introduced above modify the production and decay of the Higgs boson, particularly in processes involving neutral and charged electroweak gauge bosons. The CP even terms interfere with SM amplitudes, allowing their effects to appear at $\mathcal{O}(1/\Lambda^{2})$ in observables like cross sections and decay rates. In contrast, the CP odd contributions remains suppressed to $\mathcal{O}(1/\Lambda^4)$ to such observables making their detectability far more challenging. 

To overcome this limitation, spin related observables provide a sensitive probe to such CP violating NP. The next section describes the construction of spin related observables used in this study and the specific asymmetries employed to isolate the effects of the various SMEFT operators.
\section{Observables: CP-even and CP-odd sensitivity}
\label{sec:spin}

In the presence of both a CP-even operator $(\mathscr{O}_E)$ and a CP-odd operator $(\mathscr{O}_O)$ at dimension six, the structure of any observable can be understood from the squared matrix element,
\begin{align}
    |\mathcal{M}|^2 
    &= \left| \mathcal{M}_{\mathrm{SM}} + \mathcal{M}_{E} + \mathcal{M}_{O} \right|^2 \nonumber \\
    &= |\mathcal{M}_{\mathrm{SM}}|^2 
    + 2~\mathrm{Re}\left(
        \mathcal{M}_{\mathrm{SM}}^{\star}\mathcal{M}_{E}
        + \mathcal{M}_{\mathrm{SM}}^{\star}\mathcal{M}_{O}
        \nonumber\right.\\&+\left. \mathcal{M}_{E}^{\!*}\mathcal{M}_{O}
    \right)
    + |\mathcal{M}_{E}|^2 
    + |\mathcal{M}_{O}|^2.
    \label{eq:amp_expansion}
\end{align}
This expansion is completely general, and the contribution of each term depends on the CP properties of the observable under consideration. For CP-even observables—such as the total or differential cross section—the interference terms that are odd under a CP transformation vanish after phase–space integration. In particular, the terms
\[
    2~\mathrm{Re}\left(
        \mathcal{M}_{\mathrm{SM}}^{\star}\mathcal{M}_{O}
        + \mathcal{M}_{E}^{\star}\mathcal{M}_{O}
    \right)
\]
do not contribute, as they are CP-odd by construction. Consequently, the leading NP contribution to CP-even quantities arises from the interference between the SM amplitude and the CP-even operator, scaling as $\mathcal{O}(\Lambda^{-2})$, followed by pure dimension-6 squared contributions of order $\mathcal{O}(\Lambda^{-4})$.

In contrast, CP-odd observables are specifically designed to be odd under CP transformations. For such quantities, the interference between the SM and the CP-odd operator, $\mathrm{Re}(\mathcal{M}_{\mathrm{SM}}^{\star}\mathcal{M}_{O})$, provides the leading non-vanishing term, contributing at $\mathcal{O}(\Lambda^{-2})$. Additional effects from the interference between two dimension-6 operators, $\mathrm{Re}(\mathcal{M}_{E}^{\star}\mathcal{M}_{O})$, appear at $\mathcal{O}(\Lambda^{-4})$.

Hence, the construction of observables sensitive to CP-odd effects is essential, as they retain the interference terms that vanish in inclusive or CP-even measurements, thereby providing enhanced sensitivity to CP-violating NP already at the interference level.

The polarization and spin correlation parameters of the massive spin-1 boson provides a large set of observables which has a $\mathcal{O}(\Lambda^{-2})$ dependence with CP-violating higher dimensional operators. In general, for a system of $N$ particles with spin $s\in \{s_1,s_2,\cdots,s_n\}$, there will be a total of $4s_1(s_1+1)\times 4s_2(s_2+1) \cdots 4s_n(
s_n+1)$ spin observables.  Considering $e^-e^+\to ZH$ production, we choose three decay channels of the Higgs boson viz. $h \to b\bar{b}/WW^\star/ZZ^\star$ giving a total of $8/729/729$ different spin observables in each decay channel. For the $h \to b\bar{b}$ decay, we do not consider the reconstruction of $b$-jet polarization in the current work. The $Z$ boson polarizations were used in previous literature~\cite{Rindani:2009pb,Rao:2019hsp,Rao:2020hel,Rao:2023jpr} to probe the CP violation in Higgs sector. We study the sensitivity of each decay channel along with the combination of multiple channels.

These spin parameters are reconstructed from the joint angular distribution of the final state charged fermions. We discuss in brief the methodology of obtaining these spin observables in terms of the asymmetries in angular functions of final fermions (for detailed discussion, we refer the reader to Ref.~\cite{Boudjema:2009fz,Rahaman:2021fcz}). For a spin-1 massive boson, the production dynamics are represented in terms of density matrix as
\begin{equation}
    \rho(\lambda,\lambda^\prime) \propto \mathcal{M}(\lambda)\mathcal{M}^\star(\lambda^\prime),
\end{equation}
where $\mathcal{M}$ is the helicity amplitude and $\lambda \in [-1,0,1]$ are three possible helicity states of massive spin-1 boson. The density matrix can be parameterized in terms of three vector $\Vec{p} = \{p_x,p_y,p_z\}$ and five tensorial $T_{ij}$ polarizations, given in cartesian form as
\begin{equation}
    \rho(\lambda, \lambda^\prime) = \frac{1}{3}\left[\mathbb{I}_{3\times 3} + \frac{3}{2}\Vec{p}\cdot \Vec{S} + \sqrt{\frac{3}{2}}T_{ij}\{S_i,S_j\}\right],
\end{equation}
where $\{S_i,Sj\}$ represents the anti-commutation between the basis of spin-1 particle $S_i \in  \{S_x,S_y,S_z\}$. 

Given the spin-1 boson decay to a pair of fermions $(V \to ff^\prime)$, the differential cross~section would be given by~\cite{Boudjema:2009fz} 
\begin{align}
    \label{eq:diffone}
    \frac{1}{\sigma}\frac{d\sigma}{d\Omega_f} &= \frac{3}{8\pi}\left[\left(\frac{2}{3}-(1-3\delta)\frac{T_{zz}}{\sqrt{6}}\right) + \alpha p_z \cos\theta_f\nonumber\right.\\&+\left. \sqrt{\frac{3}{2}}(1-3\delta)T_{zz}\cos^2\theta_f\nonumber\right.\\&+\left. \left(\alpha p_x + 2\sqrt{\frac{2}{3}}(1-3\delta)T_{xz}\cos\theta_f\right)\sin\theta_f \cos\phi_f\nonumber \right.\\&+\left. \left(\alpha p_y + 2\sqrt{\frac{2}{3}}(1-3\delta)T_{yz}\cos\theta_f\right)\sin\theta_f\sin\phi_f \nonumber \right.\\&+\left. (1-3\delta)\left(\frac{T_{xx}-T_{yy}}{\sqrt{6}}\right)\sin^2\theta_f \cos(2\theta_f)\nonumber \right.\\&+ \left. \sqrt{\frac{2}{3}}(1-3\delta)T_{xy}\sin^2\theta_f \sin(2\phi_f)\right],
\end{align}
where $\sigma = \sigma_V\times \Gamma(V\to ff^\prime)$ with $\sigma_V$ the production cross~section, $\theta_f$ and $\phi_f$ are the polar and azimuth orientation of the fermion $f$, in the rest frame of the mother boson $(V)$ with its would be momentum along the $z$-axis. The initial beam direction and the mother boson momentum in the lab frame define the $x-z$ plane, i.e., $\phi=0$ plane, in the rest frame of boson as well. The expressions for spin analyzing parameter $\alpha$ and $\delta$ are given in Appendix~\ref{app:spin}. For a spin-1 boson decaying to two fermions through the vertex structure $\bar{f}\gamma^\mu \left(f_LP_L + f_RP_R\right)f^\prime V_\mu,~P_{L/R} = \frac{1}{2}(1\mp \gamma_5)$ and in the high energy limit $\delta \to 0, \alpha_W = -1$ and $\alpha_Z=-0.219$~\cite{Rahaman:2021fcz}.

The parameters of the density matrix can be obtained by performing a partial integration of differential rate with respect to polar and azimuth angle of final state fermions. For example, the asymmetries related to the $p_x$ is given by~\cite{Rahaman:2021fcz}
\begin{align}
		\label{eq:polasymm1}
        \mathcal{A}_x &= \left(\int_{\theta=0}^\pi\int_{\phi = 0}^{\pi/2}-\int_{\theta=0}^\pi\int_{\phi=\pi/2}^{3\pi/2}+\int_{\theta=0}^\pi\int_{\phi=3\pi/2}^{2\pi}\right)\nonumber\\&d\Omega \left(\frac{1}{\sigma}\frac{d\sigma}{d\Omega}\right)\nonumber\\
			&\equiv \frac{\sigma(\sin\theta\cos\phi >0)-\sigma(\sin\theta\cos\phi < 0)}{\sigma(\sin\theta\cos\phi >0)+\sigma(\sin\theta\cos\phi < 0)}\nonumber\\
			&= \frac{3}{4}\alpha_Z p_x.
	\end{align}
	The other remaining five polarization parameters can be similarly obtained through partially integrating out the differential rate. Numerically, the elements of density matrix are computed in the form of asymmetries given by
\begin{align}
    \mathcal{A}[p_i] &= \frac{\sigma(C_i > 0) - \sigma(C_i < 0)}{\sigma(C_i > 0)+\sigma(C_i < 0)} \propto p_i,~i\in\{x,y,z\},\nonumber\\
    \mathcal{A}[T_{ij}] &= \frac{\sigma(C_iC_j > 0) - \sigma(C_iC_j < 0)}{\sigma(C_iC_j > 0) + \sigma(C_iC_j < 0)}
\end{align}
where $C_x = \sin\theta\cos\phi, C_y = \sin\theta\sin\phi$ and $C_z = \cos\theta$. These angular functions or correlators relates the differential rate to polarization parameters.

Similarly to the one particle case, the two and three particle spin correlations parameters can be obtained numerically in the form of asymmetries in the joint angular functions of final fermions. Further, these asymmetries behaves differently under CP transformation. For e.g., the polarization asymmetries $\mathcal{A}_y,~\mathcal{A}_{xy},$ and $\mathcal{A}_{yz}$ are odd under CP, while other one particle polarization asymmetries are even under CP transformation. Similarly, the correlation observables are also divided into two class of asymmetries depending upon their reaction to CP transformation (see ref.~\cite{Subba:2022czw} for detailed classification). The availability of CP-odd asymmetries becomes very important to probe the CP-odd NP which otherwise would only appear in quadratically suppressed form in CP-even observables.

In the next section, we discuss the sensitivity of these observables on the WCs particularly focusing on the CP-odd operators. We will begin with the dominant $b\bar{b}$ decay channel where we focus on one particle polarization asymmetries and move on to three particle system.

\section{Sensitivities of spin asymmetries}
\label{sec:res1}
In this section, we present a detailed study of the sensitivity of various Higgs decay channels to the dimension-six operators listed in Eq.~\eqref{eq:op}, considering quadratic dependence on the corresponding WCs. The cross section for any observable process is parameterized as
\begin{equation}
    \label{eq:xsec}
    \sigma(c_i) = \sigma_0 + \sum_{i=1}^{3} \sigma_i c_i + \sum_{i \ge j} \sigma_{ij} c_i c_j + \sum_{k=1}^{6}\sigma_{kk} c_k^2,
\end{equation}
where $\sigma_0$ denotes the SM cross~section, $\sigma_i$ encodes the interference between the SM and the three CP-even dimension-six operators, $\sigma_{ij}$ represents the interference between two CP-even operators, and $\sigma_{kk}$ captures the quadratic contributions from both CP-even and CP-odd operators.  

For asymmetry observables, we parameterize the cross-section weighted asymmetries ($\mathcal{A}\sigma/\sigma$) according to their transformation properties under CP. The CP-even asymmetries are fitted using Eq.~\eqref{eq:xsec} for their numerator $\Delta\sigma = \mathcal{A}\sigma$, while the CP-odd asymmetries are parameterized as
\begin{equation}
    \label{eq:odd}
    \Delta\sigma(c_i) = \sum_{i=1}^{3} \sigma_i c_i + \sum_{i \ge j} \sigma_{ij} c_i c_j,
\end{equation}
where the linear term $\sigma_i$ arises solely from the three CP-odd operators, and $\sigma_{ij}$ corresponds to interference terms between CP-even and CP-odd contributions.
\\
\textbf{Higgs decay to $b\bar{b}$ pair:}
We begin with the dominant decay mode of the Higgs boson, $H\to b\bar{b}$ (BR $\approx 58\%$), studied in the associated production channel $Zh \to \ell^+\ell^- b\bar{b}$. In this topology the primary background arises from the irreducible $ZZ$ process, while the radiative mode $ZZ\gamma$ can also contribute. In the present analysis, however, the requirement of a minimum $p_T$ for photons emitted in the final state suppresses the $ZZ\gamma$ contamination to a negligible level. We additionally account for the impact of initial-state radiation (ISR), which can modify the effective center-of-mass energy and consequently alter selection efficiencies.

Signal and background samples are generated at matrix-element level using \textsc{MadGraph5\_aMC\@NLO}~\cite{Alwall:2011uj}, interfaced with \textsc{Pythia}~\cite{Bierlich:2022pfr} for parton showering and hadronization, and passed through a \textsc{Delphes}~\cite{deFavereau:2013fsa} simulation configured for the ILC detector.

\begin{figure}[!t]
     \centering
     \includegraphics[width=0.49\textwidth]{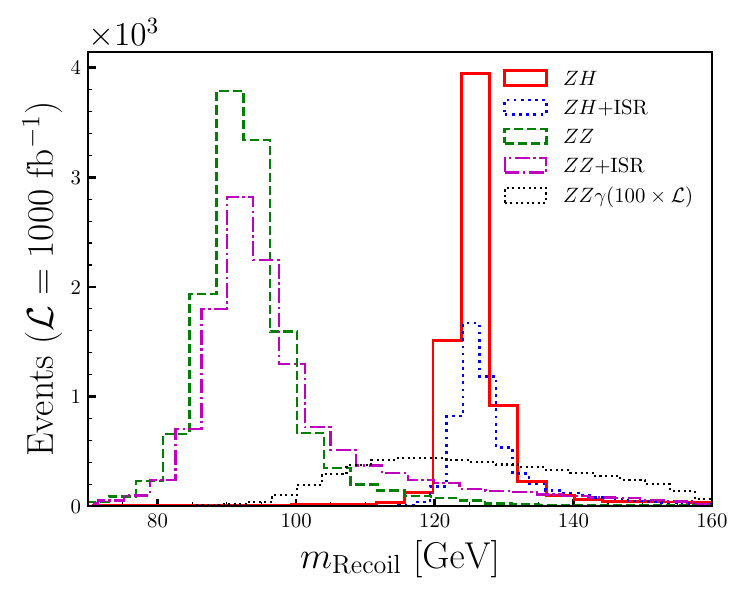}
     \caption{Recoil mass distribution normalized to $\mathcal{L}=1000$ fb$^{-1}$ for signal and two background processes viz. $ZZ$ and $ZZ\gamma$. The kinematic effect of ISR are highlighted for $Zh$ and $ZZ$ processes.}
     \label{fig:dist}
 \end{figure}

The Higgs recoil mass, reconstructed solely from the dilepton system originating from the $Z$ boson, is defined as
\begin{equation}
m_{\mathrm{Recoil}}^2 = s - 2\sqrt{s}E_{\ell^+\ell^-} + m_{\ell^+\ell^-}^2.
\end{equation}
Figure~\ref{fig:dist} shows the normalized recoil-mass distributions for the signal and the dominant backgrounds ($ZZ$ and $ZZ\gamma$) at $\mathcal{L}=1000$ fb$^{-1}$, both with and without ISR. After imposing the photon $p_T$ selection, the residual $ZZ\gamma$ contribution becomes small; for clarity of visualization its distribution is rescaled by a factor of 100. ISR reduces the event yield in the $Zh$ sample by lowering the effective interaction energy below $250$ GeV, thereby reducing the production cross section. The $ZZ$ background, in contrast, exhibits negligible distortion in shape or overall rate. Importantly, the recoil-mass observable retains strong discriminating power between signal and background even in the presence of ISR.

Events are selected by imposing $m_{\mathrm{Recoil}} \geq 120~\mathrm{GeV}$ to effectively suppress the dominant $ZZ$ background, as shown in Figure~\ref{fig:dist}. The advantage of this recoil mass reconstruction is that it remains independent of the Higgs decay channel. 
 \begin{figure*}[htb!]
    \centering
    \includegraphics[width=0.32\textwidth]{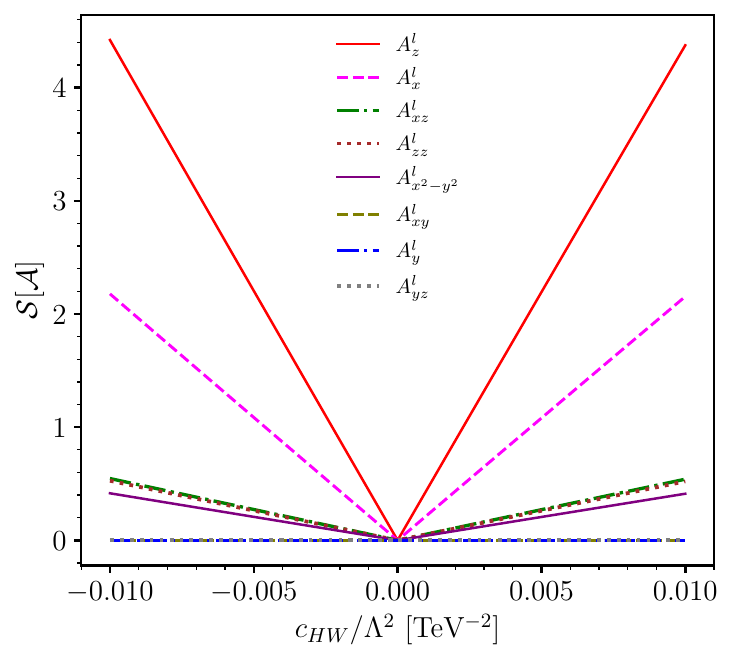}
    \includegraphics[width=0.32\textwidth]{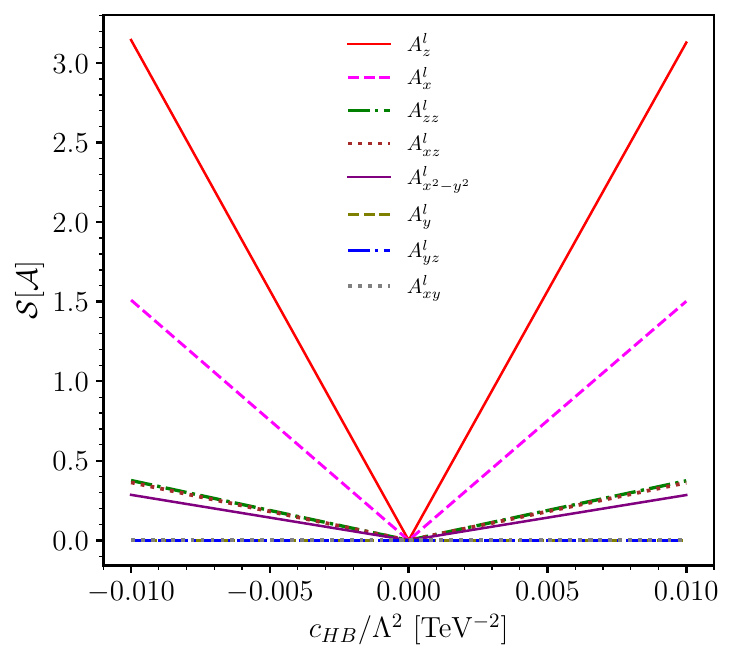}
    \includegraphics[width=0.32\textwidth]{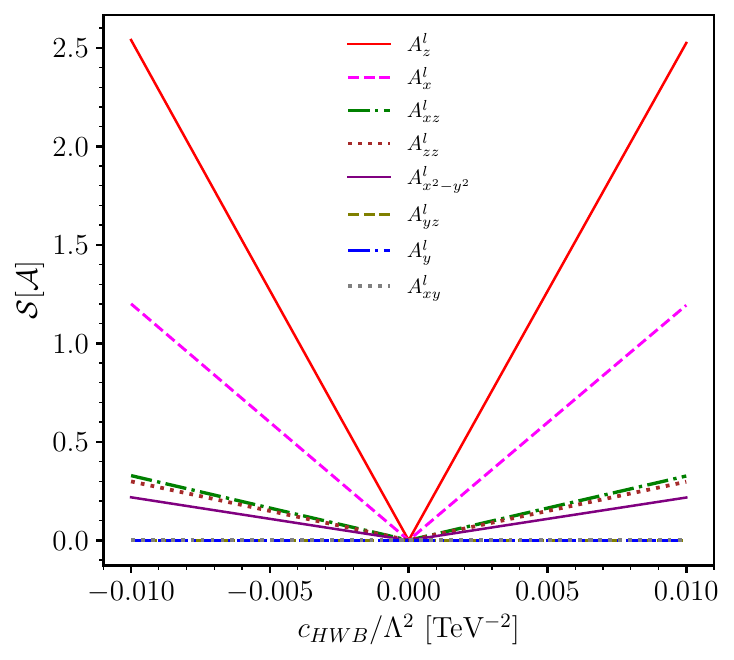}
    \includegraphics[width=0.32\textwidth]{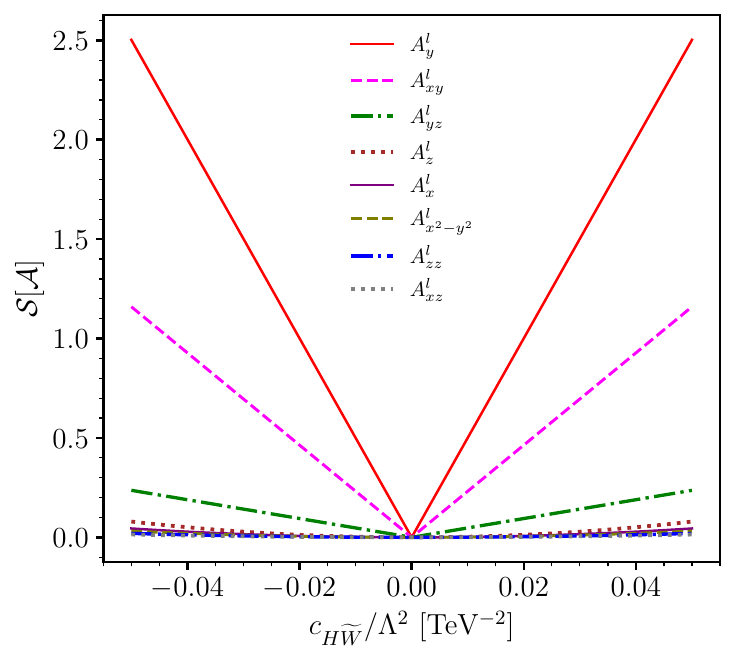}
    \includegraphics[width=0.32\textwidth]{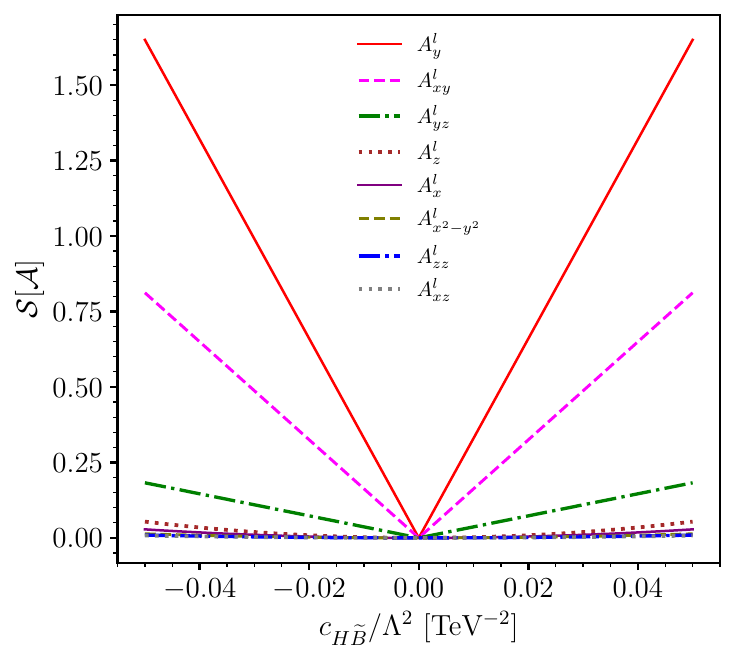}
    \includegraphics[width=0.32\textwidth]{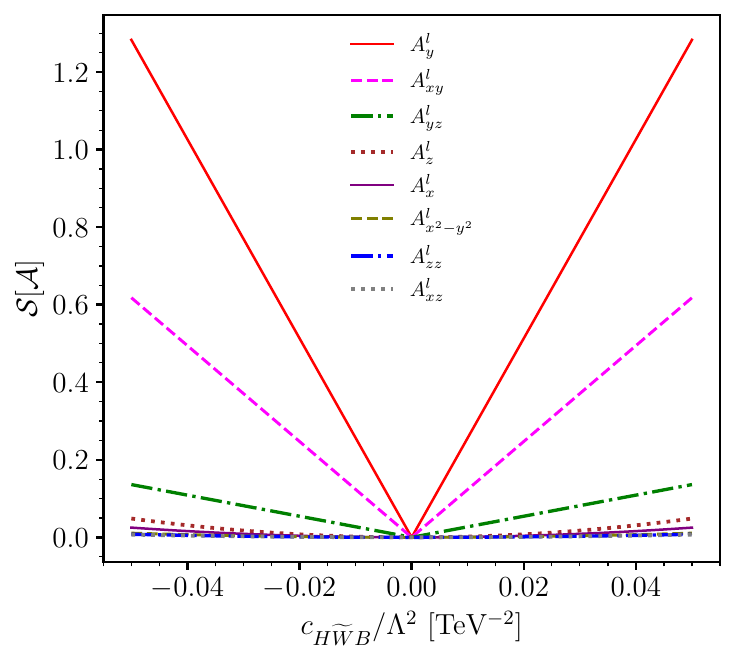}
    \caption{Sensitivity of asymmetries to the WCs of the dimension-6 operators affecting $hZZ$ and $hZ\gamma$ coupling in $Zh$ production process. The distribution are obtained for $e^-e^+ \to Zh \to l^-l^+b\bar{b}$ process at $\sqrt{s}=250$ GeV, $\Lambda=1$ TeV with initial polarized beams, $(P_{e^-},P_{e^+}) = (\mp0.8,\pm0.3)$ and an integrated luminosity of $\mathcal{L}=1$ ab$^{-1}$ for each polarization setting. No systematic errors are considered.}
    \label{fig:bbsen}
\end{figure*}

The anomalous interactions here arise from modifications to the $hZZ$ and $hZ\gamma$ vertices (Eqs.~\eqref{eq:hzz} and \eqref{eq:hza}). Since the $b$-quark polarization is not reconstructed, only the $Z$-boson spin parameters contribute to the angular structure. The angular observables are extracted from leptons in the rest frame of the decaying $Z$ boson. To enhance statistical sensitivity, events are categorized in bins of $\cos\theta_Z$, where $\theta_Z$ is the production angle of the $Z$ boson in the lab frame as
\begin{equation}
    B_n = -1 + \frac{(n-1)}{4} < \cos\theta_Z \leq -1 + \frac{n}{4}, \quad n = 1,\dots,8.
\end{equation}

The sensitivity of a given observable $\mathcal{O}$ to a Wilson coefficient $c$ is defined as 
\[
\mathcal{S}[\mathcal{O},c] = \frac{|\mathcal{O}(c) - \mathcal{O}(0)|}{\delta\mathcal{O}},
\]
where $\delta\mathcal{O} = \sqrt{\delta\mathcal{O}_{\mathrm{syst.}}^2+\delta\mathcal{O}_{\mathrm{syst.}}^2}$ denotes the estimated uncertainty in the measurement of observable $\mathcal{O}$.  
Figure~\ref{fig:bbsen} shows the resulting sensitivity distributions at $\mathcal{L} = 1~\mathrm{ab}^{-1}$ for each set of beam polarization $(P_{e^-},P_{e^+}) = (\mp0.8,\pm0.3)$.
For CP-even operators i.e., $\mathscr{O}_{HW}$, $\mathscr{O}_{HB}$, and $\mathscr{O}_{HWB}$, we observe that the most sensitive observables are the CP-even asymmetries $\mathscr{A}_z$ followed by $\mathscr{A}_x$. These asymmetries receive leading-order contributions at $\mathcal{O}(1/\Lambda^2)$ due to interference between the SM and the CP-even SMEFT amplitudes. In contrast, the CP-odd asymmetries, namely $\mathscr{A}_y$, $\mathscr{A}_{xy}$, and $\mathscr{A}_{yz}$, exhibit negligible sensitivity to these operators. This behavior is expected, as contributions from CP-even operators to CP-odd observables only arise at second order in the SMEFT expansion, i.e., $\mathcal{O}(1/\Lambda^4)$, and are therefore strongly suppressed.

On the other hand, when we consider CP-odd operators, we observe a complementary pattern. The asymmetry $\mathscr{A}_{y}$ emerges as the most sensitive observable across all three CP-odd operators studied. This is consistent with theoretical expectations, as $\mathscr{A}_{y}$ is explicitly CP-odd and can receive leading-order contributions at $\mathcal{O}(1/\Lambda^2)$. This highlights the unique role of CP-odd asymmetries in isolating contributions from CP-violating SMEFT operators, which would otherwise be challenging to probe using CP-even observables alone.
\\

\noindent\textbf{Higgs decay to $WW^\star~(h \to l\nu jj)$:}
We now turn our attention to the decay mode of the Higgs boson into a pair of $W$ bosons, specifically the semi-leptonic final state of the process $h \to W^\star W \to \ell \nu jj$. This channel constitutes the second most dominant Higgs decay mode~(BR $\approx 21\%$) at the LHC after $h \to b\bar{b}$, and provides complementary sensitivity to anomalous couplings due to the spin-1 nature of the $W$ bosons and their correlations.

The simulation pipeline follows similar to that of the $b\bar{b}$ channel with following parton level kinematic cuts,
\begin{align}
	\label{eq:cut}
        &p_T^j \ge 20~\mathrm{GeV}, \quad p_T^\ell \ge 10~\mathrm{GeV},\qquad |\eta_\ell| \le 2.5\nonumber \\ &\Delta R_{ab} \ge 0.4 \quad (a,b \in \{\ell,j\}),
        |\eta_j| \le 5.0,,
\end{align}
where $\Delta R_{ab} \equiv \sqrt{(\Delta\eta)^2 + (\Delta\phi)^2}$ denotes the angular separation between any pair of leptons or jets in the final state, and $\eta$ represents the pseudorapidity.

The event selection focuses on final states with exactly three isolated leptons, two jets, and missing transverse energy, corresponding to $Zh \to \ell^+ \ell^- + h(W^{(*)} W \to \ell \nu jj)$. A $Z$ boson candidate is reconstructed by requiring a pair of same-flavor, oppositely charged leptons with an invariant mass close to the nominal $Z$-boson mass, specifically
\[
|m_{\ell^+\ell^-} - m_Z| \le 5~\mathrm{GeV}, \quad \text{with } m_Z = 91.19~\mathrm{GeV}.
\]
In addition, to ensure that the selected events are consistent with the associated $Zh$ production topology, we require the recoil mass against the identified dilepton system form $Z$ boson to satisfy $m_{\mathrm{Recoil}} \ge 120$ GeV thereby suppressing background contributions from processes not associated with on-shell Higgs boson production.

The dimension-six operators listed in Eq.~\eqref{eq:op} modify both the production and decay dynamics of this channel. On the production side, deviations arise through anomalous contributions to the $hZZ$ and $hZ\gamma$ couplings, which affect the angular correlations and kinematic distributions of the $Zh$ system. On the decay side, the Higgs coupling to $W$ bosons, $hWW$, is altered by additional effective operators, explicitly shown in Eq.~\eqref{eq:hww}, which introduce new Lorentz structures and modify the spin correlation between the decay products. These modifications can manifest as distortions in the angular observables of the final-state particles.

The $h\to WW^\star$ decay channel would allow for two and three body spin correlations along with three spin-$1$ polarizations parameters. However, to reconstruct the vector polarization associated with the hadronic $W$ decay and their correlations, one needs to know the identity of final two jets. In the current work, we follow the tagging procedure developed in previous works~\cite{Subba:2022czw,Subba:2023rpm} using boosted decision trees (BDT) to tag the identity of final two jets in \emph{up/down}-type class. The final two jets are reconstructed using the anti-$k_T$~\cite{Cacciari:2008gp} clustering algorithm. For truth labeling of the two jets, we use angular distance $\Delta R$, between the two hardest jets and two parton level light quarks. In the case, when both the jets are closer to one type of quarks, we truth tagged the hardest jet of that particular quark flavor. Further, we construct several continuous and discrete observables from the jets to used as an input features for our BDT model. The list of features are listed in Table~\ref{tab:flavtag}.  Continuous variables, such as particle momenta, are normalized by the jet energy. We also compute the number of mother particles that has traveled a distance $d > 0.3$ mm from primary vertex. Such particles gives raise to final displaced tracks (tracks originating from a secondary vertex). And in the case of jet initiated by $c$ quark, we can have significant number of short lived kaons ($K^0_S$) leading to such displaced tracks. The $p_T$ weighted jet-charge variables are also used as an additional features. 
\begin{table}[!t]
    \centering
    \renewcommand{\arraystretch}{1.5}
    \caption{\label{tab:flavtag} List of observables used as input features for BDT in order to classify two leading jets as either \emph{up/down}-type jets. The features are derived for $e^-e^+ \to Zh \to 3l\slashed{E}2j$ process at $\sqrt{s}=250$ GeV at the detector (\textsc{ Delphes}) level.}
    \begin{tabular*}{0.49\textwidth}{@{\extracolsep{\fill}}cc@{}}
    \hline
     Feature & Description  \\
     \hline
      \( N_\gamma \), \( N_l \), \( N_{\pi^\pm} \), \( N_{k^\pm} \) & Number of $\gamma$, $l^\pm$, $\pi^\pm$, and $k^\pm$  \\
      \( p_\gamma, p_l, p_\pi, p_k \) & Four-momentum of $\gamma$, $l^\pm$, $\pi^\pm$, and $k^\pm$ \\
      \( p_T^\gamma, p_T^l, p_T^{\pi^\pm}, p_T^{k^\pm} \) & Scalar sum \( p_T \) of respective particles \\
      \( Q_J^k \) & Jet charge (\( k \in \{0,1\} \))  \\
      $N_d$ & Count of mother with lifetime $d > 0.3$ mm. \\
      \hline
    \end{tabular*}
\end{table}
\begin{figure*}[!htb]
	\centering
	\includegraphics[width=0.32\textwidth]{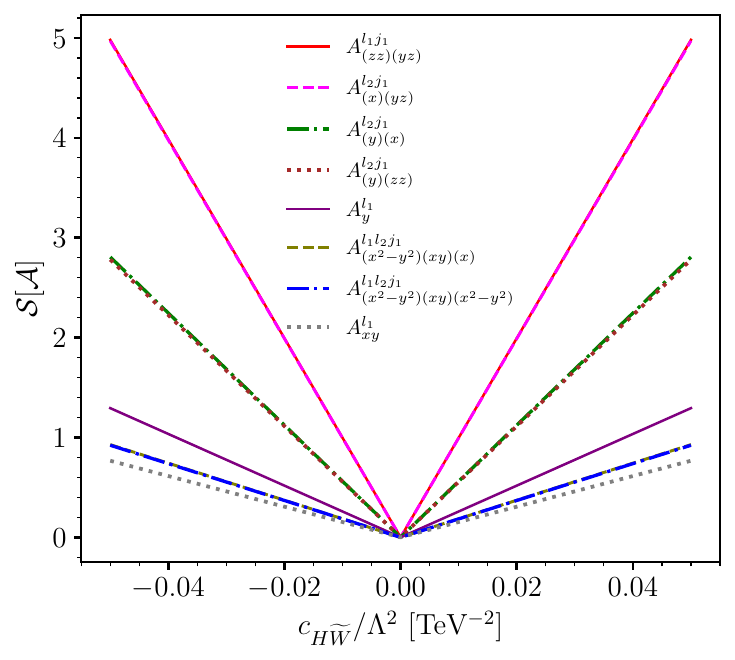}
	\includegraphics[width=0.32\textwidth]{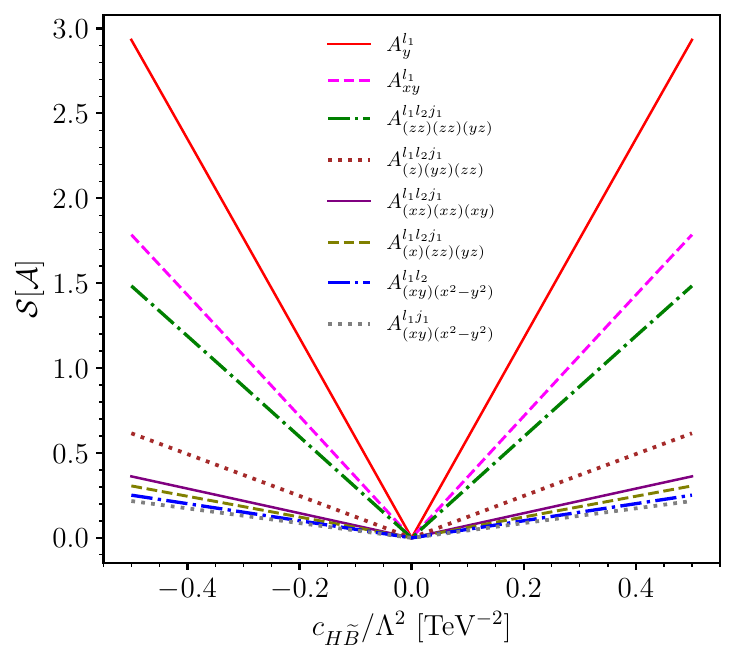}
	\includegraphics[width=0.32\textwidth]{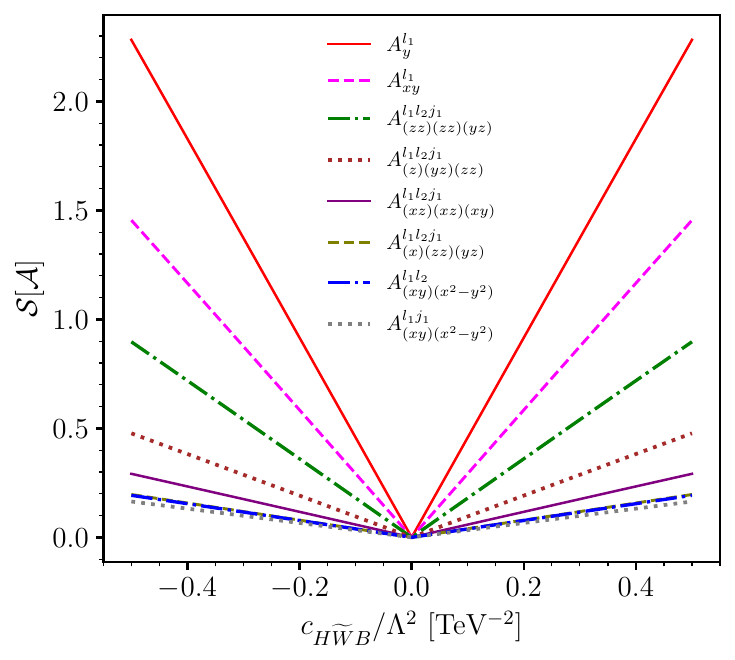}
    \includegraphics[width=0.32\textwidth]{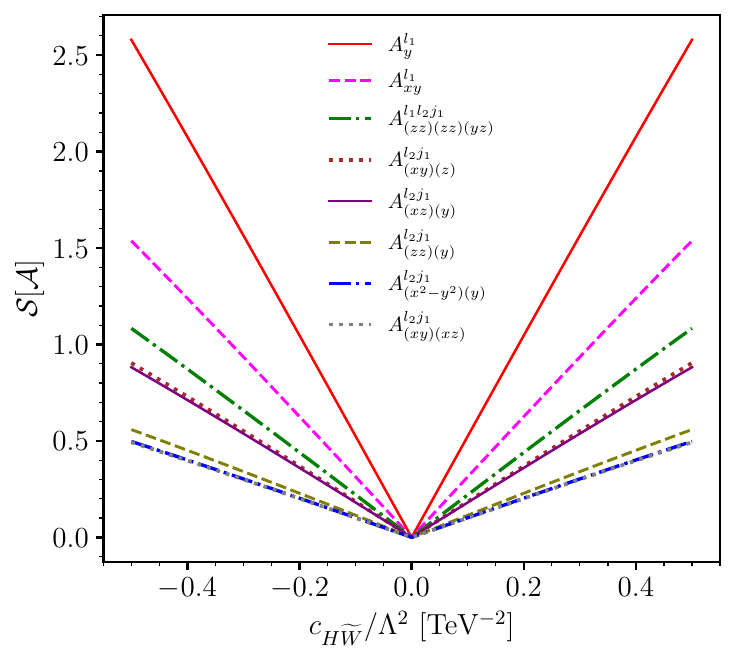}
	\includegraphics[width=0.32\textwidth]{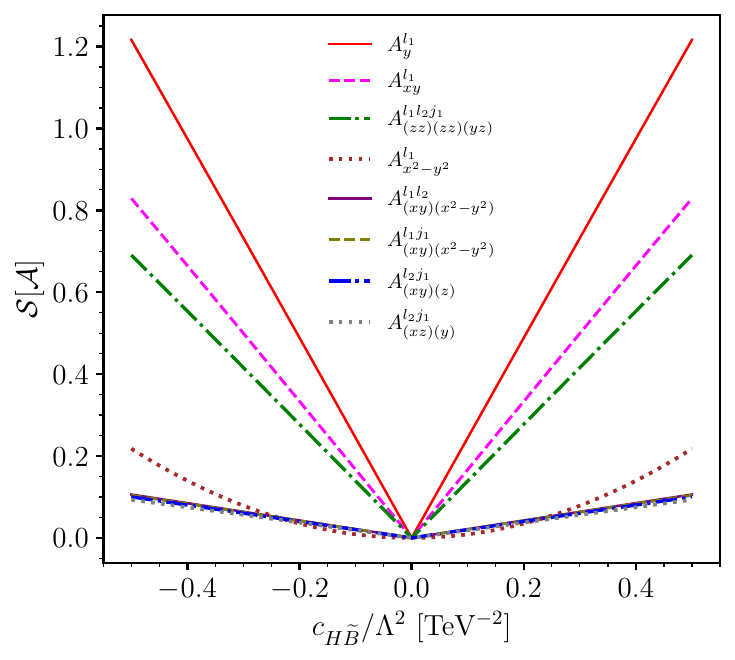}
	\includegraphics[width=0.32\textwidth]{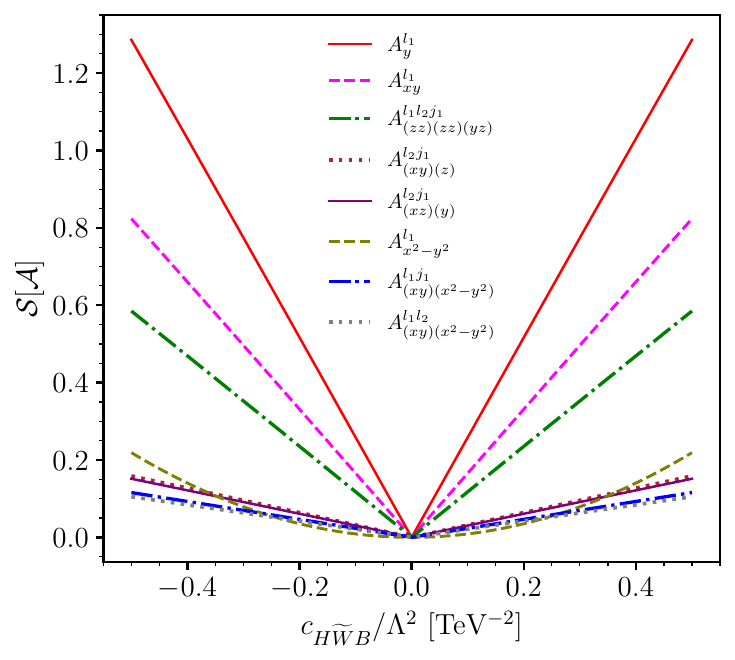}
	\caption{\label{fig:wwzzcpodd}In the top row, we show the sensitivity for asymmetries as a function of one WC at a time obtained with $h \to WW^\star$ channel and in the bottom row, the sensitivity for $h\to ZZ^\star$ are shown. Only top eight sensitive asymmetries for each CP-odd WC are shown. The distribution are obtained at $\sqrt{s}=250$ GeV, $\Lambda =1$~TeV with beam polarization $(P_{e^-},P_{e^+}) = (\mp0.8,\pm0.3)$ and an integrated luminosity $\mathcal{L}=1$ ab$^{-1}$. No systematic errors are taken in this analysis. }
\end{figure*}
The classification model is based on an \textsc{XGBoost}~\cite{Chen_2016} binary classifier with the binary logistic objective and evaluated using the log-loss metric. A grid search over learning rate, number of boosting rounds, and tree depth identified the optimal configuration: learning rate of $0.05$, $500$ boosting rounds, and a maximum depth of 10. Column and row sub-sampling $(0.7)$ and regularization terms (L1 = 1.0, L2 = 1.0) were used to control over-fitting. The model was trained using 3-fold cross-validation to ensure generalizability. Model performance was evaluated via iterative sub-sampling of test data, achieving \( \approx 78\% \) classification accuracy with a variance of \( \approx 0.01\% \). The trained model is further used to reconstruct the spin asymmetries of three spin-1 bosons.

The sensitivities of the \( h \to WW^\star \) channel to the three CP-odd dimension-6 operators are shown in top row of Figure~\ref{fig:wwzzcpodd}. The plots display the response of the most significant eight angular asymmetries to variations in \( c_{H\widetilde{W}} \), \( c_{H\widetilde{B}} \), and \( c_{H\widetilde{W}B} \). For the \( c_{H\widetilde{W}} \) coupling, the mixed asymmetry \( \mathcal{A}_{(zz)(yz)}^{l_1 j_1} \), which correlates the \(\sin(3\theta)\) mode of the lepton from the \(Z\) boson with the \(\cos\theta\sin\phi\) mode of the leading jet from the \(W\) boson, exhibits the highest sensitivity. A comparable enhancement is observed for \( \mathcal{A}_{(yz)(x)}^{l_2 j_1} \), indicating strong correlations between angular structures in the leptonic and hadronic decay sectors.

For the \( c_{H\widetilde{B}} \) and \( c_{H\widetilde{W}B} \) coefficients, the dominant sensitivity arises from the parity-odd lepton asymmetry \( \mathcal{A}_y^{l} \). This is followed by \( \mathcal{A}_{xy}^{l} \), which encodes azimuthal–polar correlations characteristic of CP-violating effects in the leptonic decay plane. Overall, the leading sensitivities are governed by single- and mixed-angular observables that transform odd under parity and CP, making them optimal probes of CP-odd interactions in the \( WW^\star \) topology. It should be noted that, although quadratic contributions from SMEFT operators are included in the analysis, Figures~\ref{fig:bbsen} and \ref{fig:wwzzcpodd} show that the resulting constraints are sufficiently stringent that these quadratic effects remain numerically insignificant.
\\

\noindent\textbf{Higgs decay to four leptons~$(h \to e^-e^+\mu^-\mu^+)$:}
We finally analyze the fully leptonic channel $H \to ZZ^{(*)} \to e^+e^-\mu^+\mu^-$, corresponding to the $Zh \to \ell^+\ell^- + h(ZZ^\star\to e^+e^-\mu^+\mu^-)$ topology. Despite its smaller branching ratio ($\approx 3\%$), this process offers clean kinematics and negligible background, enabling full reconstruction of the intermediate $Z$ polarizations.  Moreover, due to the spin-1 nature of the $Z$ bosons, this channel is highly sensitive to the tensor structure of the $hZZ$ vertex, making it a powerful probe of anomalous Higgs couplings arising from higher-dimensional operators.

At the parton level, basic kinematic acceptance cuts are applied to all leptons to ensure detector-level observability and to regulate soft and collinear divergences. These cuts are implemented in MG5 and are given by
    \begin{align}
        p_T^\ell \geq 10~\mathrm{GeV}, \qquad \Delta R_{\ell_1 \ell_2} \geq 0.4, \qquad |\eta_\ell| \leq 2.5,
    \end{align}
where $\Delta R_{\ell_1 \ell_2} = \sqrt{(\Delta \eta)^2 + (\Delta \phi)^2}$ is the angular separation between any pair of leptons, and $\eta_\ell$ denotes the pseudorapidity.

To ensure that the selected events originate from an on-shell $Z$ boson and to suppress non-resonant background, the recoil mass against the dilepton system is required to lie within the Higgs mass window:
\[
123~\mathrm{GeV} \leq m_{\mathrm{recoil}} \leq 127~\mathrm{GeV},
\]
where the recoil mass is computed as
\[
m_{\mathrm{recoil}} = \sqrt{ (p_{\mathrm{initial}} - p_{\ell^+\ell^-})^2 },
\]
with $p_{\mathrm{initial}}$ being the total incoming momentum and $p_{\ell^+\ell^-}$ the momentum of the tagged $Z$ boson.

The Higgs boson candidate is reconstructed from the remaining four-lepton final state consisting of two electrons and two muons. To ensure compatibility with the Higgs resonance, the invariant mass of the four-lepton system is required to satisfy
\[
120~\mathrm{GeV} \leq m_{4\ell} \leq 130~\mathrm{GeV}.
\]
This selection efficiently suppresses continuum $ZZ$ backgrounds and enhances the signal purity of Higgs decays.
\begin{table*}[!htb]
    \centering
    \caption{\label{tab:onecom}$95\%$ C.L. one parameter limits of WCs obtained using cross~section and asymmetries for three different decay channel of the Higgs boson and also their combinations for $e^-e^+ \to Zh$ production. The limits are obtained at $\sqrt{s}=250$ GeV, $\Lambda = 1$ TeV, $\mathcal{L}=1000$ fb$^{-1}$ for each set of beam polarization with zero systematics. }
    \renewcommand{\arraystretch}{1.3}
	  	\begin{tabular*}{\textwidth}{@{\extracolsep{\fill}}c*{4}{>{}c<{}}@{}}
	  		\hline 
            WCs & $l^-l^+b\bar{b}$ & $l^-l^+ WW^\star$ & $l^-l^+ ZZ^\star$  & Combined \\
            \hline
            $C_{HW}$ & $[-0.012,+0.012]$ & $[-0.001,+0.001]$ & $[-0.081,+0.050]$ & $[-0.001,+0.001]$ 
            \\
            $C_{H\widetilde{W}}$&$[-0.190,+0.190]$&$[-0.041,+0.041]$&$[-3.721,+3.721]$&$[-0.040,+0.040]$
            \\
            $C_{HB}$ & $[-0.020,+0.020]$&$[-0.101,+0.098]$&$[-0.775,+0.502]$&$[-0.020,+0.020]$
            \\
            $C_{H\widetilde{B}}$&$[-0.325,+0.325]$&$[-1.032,+1.032]$&$[-6.670,+6.667]$&$[-0.322,+0.312]$
            \\
            $C_{HWB}$&$[-0.015,+0.015]$&$[-0.111,+0.108]$&$[-0.508,+0.271]$&$[-0.014,+0.014]$
            \\
            $C_{H\widetilde{W}B}$&$[-0.360,+0.360]$&$[-1.148,+1.148]$&$[-6.889,+6.889]$&$[-0.346,+0.346]$
            \\
            \hline 
        \end{tabular*}
\end{table*}
The sensitivity of the \(h\to ZZ^\star \) channel to the CP-odd operators is shown in bottom row of Figure~\ref{fig:wwzzcpodd}, where we highlight eight dominant asymmetries. The distributions are rescaled by a factor of ten for better visualization, as the deviations in this channel are statistically suppressed. For all three WCs \( c_{H\widetilde{W}} \), \( c_{H\widetilde{B}} \), and \( c_{H\widetilde{W}B} \), the most sensitive observable is the parity-odd lepton asymmetry \( \mathcal{A}_y^{l_Z} \), arising from the azimuthal modulation of the lepton associated with the on-shell \( Z \) boson. The next leading sensitivities are observed in mixed-angular asymmetries such as \( \mathcal{A}_{xy}^{l_Z} \) and \( \mathcal{A}_{(zz)(yz)}^{l_1 l_2} \), which encode correlations among the leptons emitted from the Higgs decay chain \( h \to ZZ^\star \). Across all operators, the sensitivity pattern remains largely consistent, indicating that the dominant effects originate from parity-odd angular structures in the leptonic decay planes. The fully leptonic \( h\to ZZ^\star \) topology thus provides a clean environment where angular asymmetries serve as sensitive probes of CP-violating interactions in the Higgs–gauge sector. 

\section{Combined Results}
\label{sec:res2}

The projected $95\%$ confidence level (C.L.) limits on the dimension-6 operators modifying the Higgs–vector boson interactions are summarized in Table~\ref{tab:onecom}. The results are derived from the analysis of three Higgs decay channels: $h\to b\bar{b}$, $h\to WW^\star$, and $h\to ZZ^\star$, based on spin asymmetries and total cross-section information in $Zh$ production at $\sqrt{s}=250$~GeV with an integrated luminosity of $1000$~fb$^{-1}$ per polarization configuration.

Among the operators considered, the $h\to WW^\star$ decay channel provides the most stringent constraints for those directly modifying the $hWW$ vertex. Specifically, $c_{HW}$ attains its strongest bound in the $h\to WW^\star$ mode with limits of $[-0.001,+0.001]$, which is about an order of magnitude tighter than the $h \to b\bar{b}$ result and nearly two orders better than the $h \to ZZ^\star$ channel. The CP-odd operator $C_{H\widetilde{W}}$ follows a similar trend, with the $h\to WW^\star$ channel giving $[-0.041,+0.041]$, compared to $[-0.190,+0.190]$ from $h\to b\bar{b}$ and $[-3.721,+3.721]$ from $h\to ZZ^\star$. This improvement reflects the strong spin-correlation sensitivity of the semi-leptonic $h\to WW^\star$ observables to both CP-even and CP-odd tensor structures in the $hVV$ interaction.

For all other operators, the $h\to b\bar{b}$ channel yields the dominant sensitivity due to its large branching ratio and well-reconstructed final state. The limits on $c_{HB}$ and $c_{HWB}$ are $[-0.020,+0.020]$ and $[-0.015,+0.015]$, respectively, which are approximately five times stronger than the corresponding bounds from $h\to WW^\star$ and over an order of magnitude better than those from $h\to ZZ^\star$. 

In the CP-odd sector, $c_{H\widetilde{B}}$ and $c_{H\widetilde{W}B}$ remain comparatively weakly constrained in individual channels due to their suppressed interference. The $ZZ^\star$ mode gives limits of $\mathcal{O}(6)$, while the $b\bar{b}$ channel improves these to around $0.3$. The combined fit tightens the bounds further to $[-0.322,+0.312]$ and $[-0.346,+0.346]$, respectively, representing nearly a tenfold improvement over the $ZZ^\star$-only sensitivity.

Overall, Table~\ref{tab:onecom} demonstrates a clear complementarity among the decay channels: the $h\to WW^\star$ channel dominates for $c_{HW}$ and $c_{H\widetilde{W}}$ due to polarization-enhanced sensitivity, the $h\to b\bar{b}$ channel leads for all remaining operators owing to statistical precision, and the $h\to ZZ^\star$ channel, while limited by event rates, provides valuable kinematic clarity supporting the combined SMEFT constraints.

At this stage, we compare the constraints on the CP-even WCs obtained in this work with the bounds previously derived using the OOT in the $h\to b\bar{b}$ channel~\cite{Bhattacharya:2025jhs}. Using only the differential cross section, the OOT analysis at an integrated luminosity of $2~\text{ab}^{-1}$ yields sensitivities of
\begin{align}
C_{HW}/\Lambda^{2} &\in [-0.04, +0.04],
C_{HB}/\Lambda^{2} \in [-0.07, +0.07], \nonumber\\
C_{HWB}/\Lambda^{2} &\in [-0.08, +0.08].
\end{align}
The incorporation of $Z$-boson polarization asymmetries in the same framework substantially improves the discriminatory power, leading to the tightened bounds
\begin{align}
C_{HW}/\Lambda^{2} &\in [-0.01, +0.01],
C_{HB}/\Lambda^{2} \in [-0.02, +0.02], \nonumber \\
C_{HWB}/\Lambda^{2} &\in [-0.02, +0.02].
\end{align}
Hence, the inclusion of polarization asymmetries of $Z$ boson alone not only enhances the sensitivity to CP-odd operators, but also introduces an additional independent $\chi^{2}$ contribution in the CP-even sector, resulting in an overall improvement of approximately a factor of four in the corresponding limits.
\begin{figure*}[!htb]
    \centering
    \includegraphics[width=0.49\textwidth]{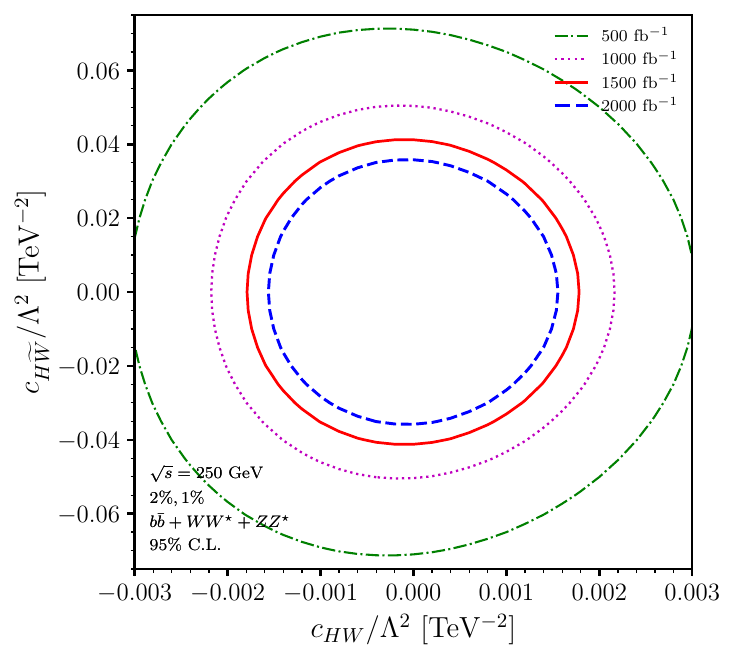}
    \includegraphics[width=0.49\textwidth]{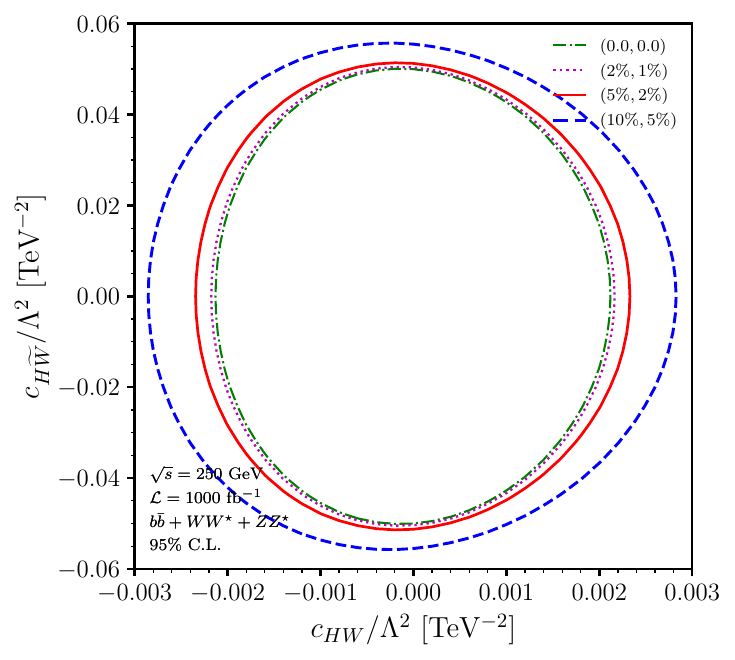}
    \caption{Impact of integrated luminosity~(left panel) and systematic uncertainties (right panel) on two-dimensional 95\% CL contours in the Wilson coefficient plane, using combined Higgs decay channels ($h\to b\bar{b}$, $h \to WW^\star$, $h\to ZZ^\star$) from $Zh$ production at $\sqrt{s} = 250$~GeV $e^-e^+$ collider.}
    \label{fig:lumisyst}
\end{figure*}

Next, we perform two-dimensional projections in the $(c_{HW},c_{H\widetilde{W}})$ plane to study the impact of integrated luminosity and systematic uncertainties on the attainable limits of the Wilson coefficients. The left panel of Figure~\ref{fig:lumisyst} illustrates the evolution of the allowed parameter space with increasing $\mathcal{L}$, assuming systematic uncertainties of $2\%$ on the total cross section and $1\%$ on asymmetries. Contours are displayed for four benchmark luminosities: 500, 1000, 1500, and 2000~fb$^{-1}$, evaluated separately for each beam polarization configuration. A clear tightening of the contours is observed with increasing luminosity, reflecting the enhanced statistical power to resolve small deviations from the SM expectations. This behavior is particularly pronounced for operators whose interference terms scale linearly with event yields, demonstrating the importance of high-statistics datasets in constraining dimension-six effects.

The right panel of Figure~\ref{fig:lumisyst} examines the role of experimental systematic uncertainties at a fixed integrated luminosity of 1000~fb$^{-1}$. Four representative uncertainty configurations are considered for the total cross section and asymmetry measurements: $(0,0)$, $(2\%,1\%)$, $(5\%,2\%)$, and $(10\%,5\%)$. As expected, increasing systematic uncertainties lead to a noticeable degradation in sensitivity, manifested as enlarged allowed regions in the $(c_{HW},c_{H\widetilde{W}})$ plane. A significant tightening of the contours is observed when moving from the conservative $(10\%,5\%)$ to the moderate $(5\%,2\%)$ scenario, indicating a strong sensitivity improvement once systematics are reduced below a few percent. Beyond this point, however, the contour areas show minimal further shrinkage, signaling a saturation in the achievable precision limited by the remaining systematic floor. This suggests that to probe the anomalous $hVV$ couplings in Higgsstrahlung processes—especially when incorporating spin-sensitive observables across multiple Higgs decay channels—both high-statistics datasets and tightly controlled systematics are indispensable. Future high-luminosity $e^-e^+$ colliders thus offer an ideal environment, where reduced experimental systematics and large integrated luminosities can jointly maximize the reach of SMEFT analyses.

\section{Summary and Conclusions}
\label{sec:con}
The precise determination of the Higgs-Gauge couplings ($hVV$, with $V = Z, W, \gamma$) remains a central goal of future high-energy $e^-e^+$ colliders, offering powerful tests of electroweak symmetry breaking and sensitivity to new physics. This study investigates anomalous $hVV$ interactions through the process $e^-e^+ \to Zh$ at $\sqrt{s}=250$~GeV, employing spin-based asymmetries that access interference effects absent in total rate measurements. These observables enable direct discrimination between CP-even and CP-odd operator contributions and significantly extend the sensitivity beyond that of previous rate-based analyses.  

By analyzing the three major Higgs decay channels: $h\to b\bar{b}, WW^{(*)}$, and $ZZ^{(*)}$-we observe a clear complementarity among them. The $h \to WW^\star$ channel provides the strongest constraints on both the CP-even and CP-odd operators $\mathcal{O}_{HW}$ and $\mathcal{O}_{H\widetilde{W}}$ that modify the $hWW$ vertex directly, driven by its rich spin correlation structure. The $h \to b\bar{b}$ mode, with its large event yield and clean reconstruction, dominates the sensitivity to operators affecting the $hZZ$ and $h\gamma Z$ interactions, while the $h \to ZZ^\star$ channel, though statistically limited, offers a clean probe of parity-odd angular structures and serves as a consistency test. A combined analysis across all decay modes yields the most stringent overall bounds, substantially improving upon both current LHC limits and previous ILC projections, particularly in the CP-odd sector.

A comparison with the recent analysis of Ref.~\cite{Atwal:2025yhw}, where constraints on the CP-violating WCs were derived using the azimuthal angle observable $\Delta\Phi_{\ell\ell}$ in the $h\to b\bar{b}^\ast$ channel at $\mathcal{L}=3~\text{ab}^{-1}$, is particularly instructive. Their study reports
\begin{align}
C_{H\widetilde{W}}/\Lambda^{2} &\in [-0.28, +0.28], 
C_{H\widetilde{B}}/\Lambda^{2} \in [-0.35, +0.35], \nonumber \\
C_{H\widetilde{W}B}/\Lambda^{2} &\in [-0.38, +0.38].
\end{align}
In contrast, our analysis of the same $h\to b\bar{b}^\ast$ decay mode yields significantly stronger bounds,
\begin{align}
C_{H\widetilde{W}}/\Lambda^{2} &\in [-0.15, +0.15], 
C_{H\widetilde{B}}/\Lambda^{2} \in [-0.26, +0.26], \nonumber \\
C_{H\widetilde{W}B}/\Lambda^{2} &\in [-0.29, +0.29].
\end{align}
Thus, for the $C_{H\widetilde{W}}$ operator in the $h\to b\bar{b}$ channel alone, our constraint is tighter by nearly a factor of three. Moreover, incorporating the $h\to W W^\ast$ decay mode substantially enhances the sensitivity: the $C_{H\widetilde{W}}/\Lambda^{2}$ coefficient is constrained to $ \in [-0.03,, +0.03]$,
representing an improvement of $\mathcal{O}(1)$ compared to the $b\bar{b}$ channel. For the remaining operators, $C_{H\widetilde{B}}/\Lambda^{2}$ and $C_{H\widetilde{W}B}/\Lambda^{2}$, our limits improve upon Ref.~\cite{Atwal:2025yhw} by approximate factors of $1.34$ and $1.31$, respectively. These bounds become even more stringent once the additional set of decay channels are included.

The impact of luminosity and systematic uncertainties has also been quantified. Increasing integrated luminosity leads to a progressive tightening of the allowed regions in the WCs planes, reflecting enhanced statistical precision. Conversely, systematic effects on total cross sections and asymmetry measurements become dominant once uncertainties exceed a few percent. Beyond this threshold, sensitivity improvements saturate, underscoring the need for both large datasets and sub-percent-level systematic control to fully exploit the precision potential of future colliders.

Overall, this work demonstrates that spin asymmetry observables substantially enhance the reach of SMEFT studies in Higgsstrahlung. They provide a clean and direct handle on CP-even and CP-odd effects, leveraging interference-driven sensitivities inaccessible to total rate analyses. When combined with multiple Higgs decay channels and polarized beams, these observables form a powerful framework for probing higher-dimensional operators with unprecedented precision. Future high-luminosity $e^-e^+$ colliders thus present an ideal environment for advancing the precision frontier of Higgs physics and testing the SM at the sub-percent level.

\section*{Acknowledgments}
	A Subba and S Bhattacharya acknowledges the support of ANRF grant CRG/2023/000580.

\appendix
\section{Spin analyzing power for $V\to f_1f_2$}
\label{app:spin}
Considering the decay of spin-1 boson ($W/Z$) to two fermions via. vertex of the form $\bar{f}_1\gamma^\mu(C_LP_L+C_RP_R)f_2 V_\mu$ with real $C_{L,R}$, the expression for spin analyzing parameter $\alpha$ and $\delta$ are given by~\cite{Boudjema:2009fz}
	\begin{align}
		&\alpha =\nonumber\\ &\frac{2(C_R^2-C_L^2)\sqrt{1+(x_1^2-x_2^2)^2-2(x_1^2+x_2^2)}}{12C_LC_Rx_1x_2 + (C_R^2+C_L^2)[2-(x_1^2-x_2^2)^2+(x_1^2+x_2^2)]},\nonumber\\
		&\delta =\nonumber\\ &\frac{4C_LC_Rx_1x_2+(C_R^2+C_L^2)[(x_1^2+x_2^2)-(x_1^2-x_2^2)^2]}{12C_LC_Rx_1x_2+(C_R^2+C_L^2)[2-(x_1^2-x_2^2)^2+(x_1^2+x_2^2)]},
	\end{align} 
	where $x_i = m_i/m$, $m_i$ is the mass of final fermion and $m$ is the mass of mother boson. At the high energy limit, the final fermions decayed from the $W/Z$ are taken to be massless, $x_1\to 0, z_2 \to 0$, then one obtains $\alpha \to (C_R^2-C_L^2)/(C_R^2+C_L^2)$ and $\delta \to 0$. Further, within the SM, in the decay of $W$, we have $C_R = 0$, hence $\alpha = -1$.

\bibliography{refer}
\end{document}